\documentclass[letter]{aa}  

\usepackage{graphicx}
\usepackage{txfonts}
\usepackage[]{hyperref}
\begin{document}

   \title{Probing outflow physics through CH$_3$CN and CH$_3$OH chemistry}



   \author{
            L. Giani\inst{1}\thanks{\email{lisa.giani@inaf.it}}
            \and
          D. Zaremba\inst{2}
          \and
          M. De Simone\inst{1,3}
          \and 
          L. Tychoniec\inst{4}
          \and 
          L. Podio\inst{1}
          \and 
          C. Codella\inst{1}
          \and
          A. Somigliana\inst{5}
            }

   \institute{
            INAF, Osservatorio Astrofisico di Arcetri, Largo E. Fermi 5, 50125, Firenze, Italy
         \and
            Department of Physics and Astronomy, University of Victoria, PO Box 3055, STN CSC, Victoria BC V8W 3P6, Canada
        \and
            European Southern Observatory, Karl-Schwarzschild-Strasse 2, D-85748 Garching bei München, Germany
        \and
            Leiden Observatory, Leiden University, PO Box 9513, 2300 RA Leiden, The Netherlands
        \and
            Max-Planck-Institut für Astronomie (MPIA), Königstuhl 17, 69117, Heidelberg, Germany           
             }

   \date{}


\abstract{
Chemical correlations between molecules provide powerful diagnostics to probe the physical conditions of protostellar outflows. In particular, the relationship between methanol (CH$_3$OH) and methyl cyanide (CH$_3$CN) offers a promising tool to investigate the chemistry and irradiation environment of shocked gas.
In this Letter, we use the CH$_3$OH/CH$_3$CN abundance ratio to constrain the physical properties of the outflow driven by the Class 0 protostar S68N using ALMA Band 3 and Band 6 observations.
Assuming local thermodynamic equilibrium (LTE), we derive excitation temperatures of 50–60 K and column densities of 2–3$\times$10$^{13}$ cm$^{-2}$ for CH$_3$CN and 3–5$\times$10$^{15}$ cm$^{-2}$ for CH$_3$OH.
The resulting CH$_3$OH/CH$_3$CN abundance ratio is nearly constant along the outflow, with values of $\sim$100–200, similar to those found in other protostellar environments. 
Using an up-to-date astrochemical model, we test whether gas-phase formation of CH$_3$CN can account for the observed ratios. We find that they are reproduced only by assuming enhanced cosmic-ray ionization rates $\zeta_{\rm CR}$ up to $\sim$10$^{-14}$ s$^{-1}$.
These results suggest that the CH$_3$OH–CH$_3$CN correlation can be used as a probe of the irradiation conditions in protostellar outflows. Further studies are required to explore the possible contribution of grain-surface formation of CH$_3$CN which could lead to a lower $\zeta_{\rm CR}$ and to extend the analysis to a larger sample of sources. 
}
   \keywords{Astrochemistry --
                ISM: molecules --
                ISM: jets and outflows -- Individual: S68N
               }

   \maketitle
%
\section{Introduction}
Understanding the chemical composition of star- and planet-forming regions is essential to reconstruct early stages stellar evolution and planetary system formation \citep[e.g.,][]{Ceccarelli2023}. Young protostars undergo intense accretion and mass-ejection, launching collimated jets and outflows whose shocks release molecules from dust‑grain mantles, enriching the surrounding gas \citep{bachiller1996}. These outflows provide ideal laboratories to investigate the interplay between physical processes and chemistry \citep{codella2017,desimone2020}.

Observations of hot cores/corinos 
and outer envelopes have revealed a correlation between methyl cyanide (CH$_3$CN) and methanol (CH$_3$OH), with a typical CH$_3$OH/CH$_3$CN ratio of $\sim$100 \citep{Bergner2017,Belloche2020,Yang2021,Chahine2022,Nazari2022,vanthoff2024}. Since CH$_3$OH forms exclusively on dust grains \citep[e.g.,][]{watanabe2002,rimola2014}, this correlation suggested a shared grain‑surface origin. However, chemical models adopting revised reaction networks \citep{giani2023revised} show that CH$_3$CN can also form in the gas phase: in hot corinos, thermally released methanol is destroyed by H$_3^+$, producing CH$_3^+$, a key precursor for CH$_3$CN. In outflows, methanol is  injected into the gas phase via sputtering \citep{flower2010}, potentially linking its abundance to CH$_3$CN.
Interferometric observations of both species are currently limited to the OMC2-FIR6c-a and L1157-B1 outflows \citep{Benedettini2007,codella2009,bouvier2025}. For FIR6c-a \citep{bouvier2025}, the observed ratios (24–1000) require enhanced cosmic-ray ionization rates ($\zeta_{\rm CR}$$\sim$10$^{-14}$ s$^{-1}$), consistent with a strongly irradiated environment \citep{ceccarelli2014-CR,Fontani2017,favre2018-CR,2023sabatini,2025redaelli}. However, the scarcity of sources and uncertainties prevent a robust assessment of whether the CH$_3$OH--CH$_3$CN link observed in hot corinos holds in shocks and if it can reliably constrain outflow physics.

In this Letter, we investigate the CH$_3$OH/CH$_3$CN ratio in the outflow of the Class 0 protostar S68N (d=445 pc; \citealt{ortiz-leon2017O,herczeg2019,Zucker2019,Zucker2020}), which is traced by several iCOMs
\citep{Tychoniec2019,Podio2021,Tychoniec2021}. 
Leveraging high-resolution ALMA observations and astrochemical modelling, we exploit the CH$_3$OH/CH$_3$CN ratio to constrain the physical and irradiation conditions along the S68N outflows.

\begin{figure*}
    \centering
    \includegraphics[width=0.365\linewidth]{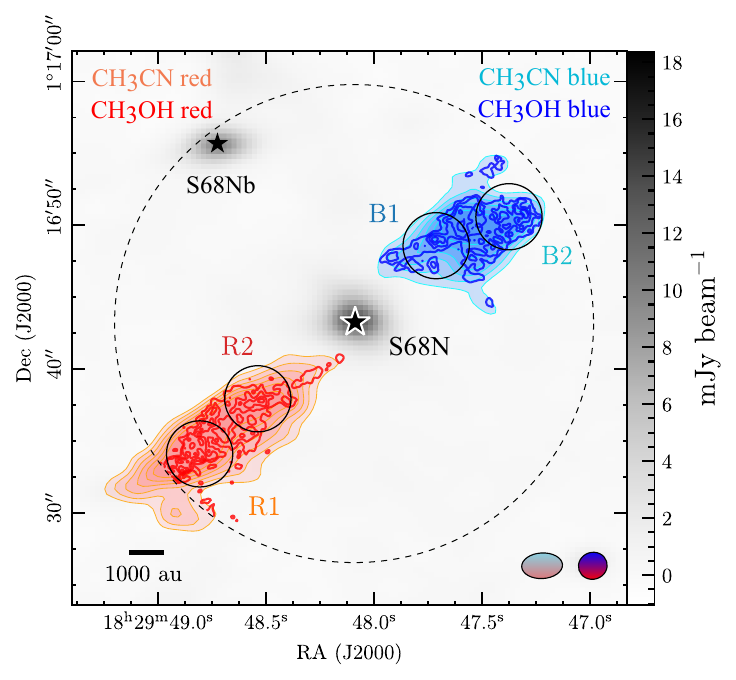}
    \includegraphics[width=0.55\linewidth]{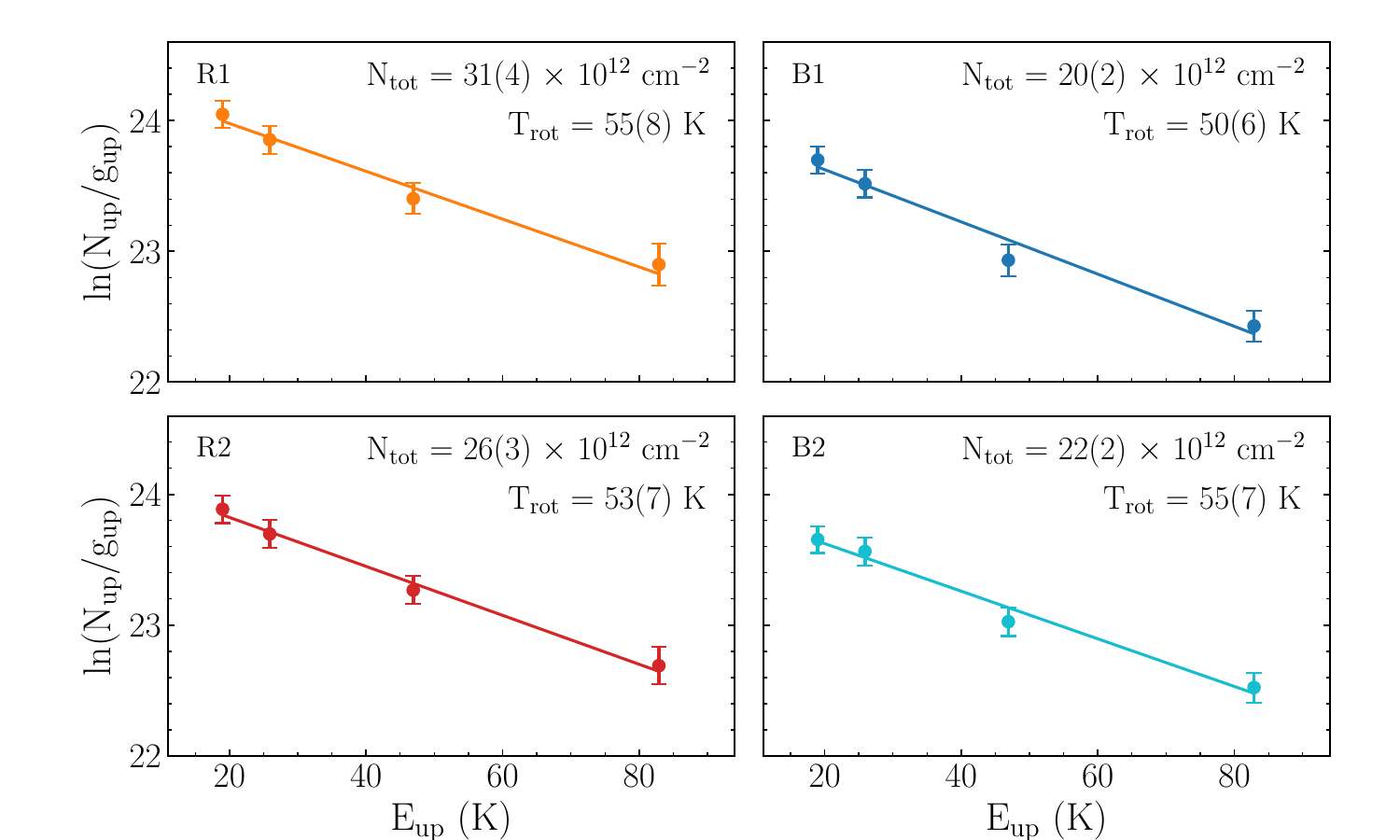}
    \caption{
    The S68N outflows and its analysis.
    \textit{Left panel:} 2.7 mm continuum (gray scale) with overlaid redshifted and blueshifted emission of CH$_3$CN $6_2-5_2$ (salmon/cyan shaded contours) and CH$_3$OH $2_{-1,1}-1_{0,1}$ (red/blue contours). 
    Contours start at 2$\sigma$ in steps of 1$\sigma$ ($\sigma$ = 60 and 50 mJy beam$^{-1}$ km s$^{-1}$ for CH$_3$CN and  CH$_3$OH, respectively). 
    Black circles label the analysed regions (R1, R2, B1, B2); stars mark protostellar positions. The dashed circle indicates the CH$_3$OH primary beam and synthesized beams are shown at the bottom right.
    \textit{Right panel:} CH$_3$CN rotation diagrams in the four labelled regions. Color coding matches the left panel. Derived column densities and rotational temperatures are reported in each panel.
    }
    \label{fig:maps+RD}
\end{figure*}

\section{Observations}\label{sec:observations}

We report high-resolution ALMA observations (2017.1.1174.S; PI: E.F. van Dishoeck) of the S68N outflows, whose details are in \citet{Tychoniec2019, vanGelder2020,Tychoniec2021}.
We present observations of CH$_3$OH $2_{-1,1}-1_{0,1}$ in band 6 and CH$_3$CN $6_0-5_0$, $6_1-5_1$, $6_2-5_2$ and $6_3-5_3$ in band 3  (Tab.~\ref{Tab:lines}). 
We used the self-calibrated data to create primary-beam-corrected images, with synthesized beams of $\sim$3$\arcsec$ ($\sim$1300 au) and $\sim$0.5$\arcsec$ ($\sim$220 au), 
spectral resolution of $\sim$0.06 MHz (0.21 km s$^{-1}$) and 0.12 MHz (0.15 km s$^{-1}$), 
field-of-views of 63$\arcsec$ and 28$\arcsec$,
and maximum recoverable scale (MRS) of 16$\arcsec$ and 6$\arcsec$
in band 3 and band 6, respectively. 
We estimated the methanol filtering-out correction to be  $\leq$20\% H$_2$CO from archival data (2013.1.00726.S; PI: C. Hull) available in two configurations with MRS of 5$\arcsec$ and 12$\arcsec$, that shows similar morphology to the CH$_3$OH emission \citep{Tychoniec2021}.

\section{Morphology and spectral analysis}\label{sec:results}

Figure~\ref{fig:maps+RD} shows dust continuum emission of S68N overlaid with CH$_3$CN and CH$_3$OH emission in the outflow. The redshifted emission is integrated between +5 and +20 km s$^{-1}$, while the blueshifted emission between --2 and 11 km s$^{-1}$. 
The systemic velocity is +8.5 km s$^{-1}$ \citep{Lee2014}.
CH$_3$CN and CH$_3$OH show a similar spatial distribution, with methanol being slightly more compact than methyl cyanide due to the lower spatial resolution. The line profiles are nearly identical, with no evidence of distinct velocity components between the two species (Fig.~\ref{fig:spectra-ch3oh-ch3cn-CO}), supporting that CH$_3$CN and CH$_3$OH trace the same gas.

\begin{figure}
    \centering
    \includegraphics[width=0.85\linewidth]{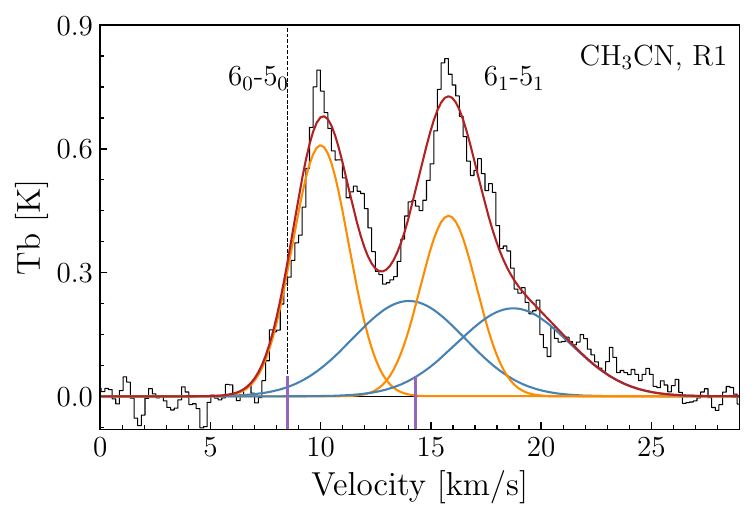}
    \caption{Spectral fits of CH$_3$CN 6$_0$–5$_0$ and 6$_1$–5$_1$ in R1 (all regions in Fig.~\ref{fig:spectra-ch3cn-deblending}).
    Orange and blue curves represent the main peaks (at $v_{\rm sys}$ $\pm$ 1 km s$^{-1}$) and the redshifted (+3-4 km s$^{-1}$ from $v_{\rm sys}$) components, respectively; red curves show the total fit. The dashed vertical line marks the systemic velocity, +8.5 km s$^{-1}$ (\citealt{Lee2014}). 
    Purple ticks indicate transition frequencies from Table~\ref{Tab:lines}.}
    \label{fig:deblending-R1}
\end{figure}

To analyse variations along the outflow, we extracted the spectra in four selected regions with a size of 2$\farcs$3 and centred on the CH$_3$CN emission peaks (Fig.~\ref{fig:maps+RD}). 
All lines show asymmetric profiles, consisting of a main component close to the systemic velocity ($\sim$6-9 km s$^{-1}$) and a secondary component shifted up to $\pm$7 km s$^{-1}$ 
(Fig.~\ref{fig:spectra-ch3oh-ch3cn-CO}). We therefore adopted a multi-component Gaussian fitting procedure to derive the total line emission. In all regions, two of the CH$_3$CN transitions ($6_0-5_0$ and $6_1-5_1$) are blended. In these cases, we used the line widths and velocities derived from the two unblended transitions to constrain the fit and separate the contributions of the blended lines (Fig.~\ref{fig:deblending-R1}). 
The fits of all lines are shown in Fig.~\ref{fig:spectra-ch3cn-deblending}.
The integrated intensities of the whole emission are reported in Tab.~\ref{Tab:lines}, while those of the low- and high-velocity components separated are reported in Tab.~\ref{Tab:lines-high-low}. 
Column densities were derived assuming local thermodynamic equilibrium (LTE) and optically thin emission, following the formalism described in \cite{Mangum2015}. Figure~\ref{fig:maps+RD} shows the rotation diagram (RD) obtained for CH$_3$CN. The temperatures derived from CH$_3$CN were then used to estimate the CH$_3$OH column densities, corrected for the filtering-out (see Sec~\ref{sec:observations}). The resulting temperatures and column densities are reported in Table~\ref{Tab:lines}. We also performed the RD analysis separately for the low- and high-velocity components (Table~\ref{Tab:lines-high-low}) to verify the presence of significant variations in temperature and CH$_3$OH/CH$_3$CN ratios among the two components.
No significant variations are found within the uncertainties; therefore, we adopt the temperatures and column densities from the total emission, which provides higher S/N and more robust estimates.
To verify the LTE conditions, we calculated the critical densities of the detected lines for CH$_3$CN and CH$_3$OH using the Einstein and collisional coefficients reported by \citet{BenKhalifa2023}, \citet{Rabli2010} and the \href{https://home.strw.leidenuniv.nl/~moldata/}{LAMDA} database.
We found n$_{\rm crit}$ $\sim$1-2 $\times$ 10$^6$ cm$^{-3}$ for CH$_3$CN and n$_{\rm crit}$ $\sim$7$\times$ 10$^7$ cm$^{-3}$ for CH$_3$OH at 50-60 K. 
To assess the validity of the LTE assumption, we performed a non‑LTE analysis of CH$_3$CN using an large velocity gradient (LVG) approach (see Appendix \ref{app-sec:LVG} for details). The results are fully consistent with those obtained from the rotational diagram analysis and indicate that the LTE approximation is appropriate: the derived gas densities are higher than $\gtrsim$10$^7$, and all CH$_3$CN transitions are optically thin.
In contrast, an LVG analysis of methanol was not feasible because only a single transition was detected. Under the assumption that CH$_3$CN and CH$_3$OH trace the same gas component, the density derived from CH$_3$CN is comparable to or higher than the critical density of methanol, suggesting that LTE conditions for CH$_3$OH are at least marginally satisfied. Furthermore, at the LTE column density derived for methanol, the optical depth remains below unity.

We derived LTE CH$_3$CN column densities ranging from 1.8 to 3.5 $\times$ 10$^{13}$ cm$^{-2}$, with gas temperatures between 44 and 68 K. 
For CH$_3$OH, we derived column densities between 3.1 and 5.3 $\times$ 10$^{15}$ cm$^{-2}$ assuming the same temperature range of CH$_3$CN. 
The CH$_3$OH/CH$_3$CN ratios calculated in the four positions range between 117 and 247, showing no significant variation with the distance from the protostar (Tab.~\ref{Tab:lines}). 
The derived ratios are in agreement with values found in hot corinos ($\sim$100; \citealt{Belloche2020,Yang2021,vanthoff2024}). The analysis of other shocked regions (L1157-B1 and OMC2-FIR6c-a; \citealt{codella2009,bouvier2025}) are not well constrained with values between 24 and 1000. Despite being qualitatively in agreement, a comparison is not constraining.

In order to estimate the abundances of the species with respect to H$_2$, we used the CO 2-1 line (230.5328 GHz, E$_{\rm up}$=17 K)
to derive the column density of CO and by consequence that of H$_2$ assuming the standard CO/H$_2$ ratio ($\sim$2$\times$10$^{-4}$; \citealt{Lacy1994}). 
In all regions (see spectra in Fig.~\ref{fig:spectra-ch3oh-ch3cn-CO}) the emission is strongly self-absorbed at systemic velocity.
At high velocities, the CO line profiles differ from those of CH$_3$OH and CH$_3$CN in all regions except R2 (v$>$12 km s$^{-1}$), making it difficult to isolate CO emission tracing the same gas component. In R2, however, the profile similarity allows a more reliable comparison. 
Thus, assuming the same temperature as CH$_3$CN and taking into account that CO emission can be optically thick, we derive an H$_2$ column density of $\geq$5$\times$10$^{20}$ cm$^{-2}$.
Using the CH$_3$OH column density from the high-velocity component (Tab.~\ref{Tab:lines-high-low}) we obtain a methanol abundance of $\leq$4$\times$10$^{-6}$.

\begin{table*}[]
    \label{tab:line-list}
    \caption{Spectroscopic parameters and line fit results over the whole emission (high + low components) of CH$_3$CN and CH$_3$OH. \label{Tab:lines}}
    \centering
    \begin{tabular}{ccccccccc}
    \hline
    \hline
    Frequency & Transition & E$_{\rm up}$ &  log(A$_{\rm ij}/{\rm s}^{-1}$) & g$_{\rm up}$ & I$_{\rm int}$ [R1] & I$_{\rm int}$ [R2] & I$_{\rm int}$ [B1] & I$_{\rm int}$ [B2]\\
    (GHz) & & (K)  &  & & (K\ km\ s$^{-1}$) & (K\ km\ s$^{-1}$)  & (K\ km\ s$^{-1}$) & (K\ km\  s$^{-1}$) \\ 
    \hline
    \multicolumn{9}{c}{CH$_3$CN}\\
    \hline
    110.3835 & $6_0-5_0$ A  & 18.5   & --3.954  & 26 & 3.43 (0.11) & 2.94 (0.11) & 2.43 (0.06) & 2.29 (0.06) \\
    110.3814 & $6_1-5_1$ E  & 25.7   & --3.966  & 26 & 2.77 (0.09) & 2.36 (0.09) & 2.02 (0.06) & 2.11 (0.07) \\
    110.3750 & $6_2-5_2$ E  & 47.1   & --4.005  & 26 & 1.62 (0.07) & 1.36 (0.05) & 1.00 (0.07) & 1.09 (0.05) \\
    110.3644 & $6_3-5_3$ A  & 82.8   & --4.079  & 26 & 1.65 (0.18) & 1.28 (0.13) & 1.03 (0.06) & 1.07 (0.06) \\
    \hline    
    N (cm$^{-2}$) & - & - & - & - & 3.1 (0.4) $\times$ 10$^{13}$ & 2.6 (0.3) $\times$ 10$^{13}$ & 2.0 (0.2) $\times$ 10$^{13}$ & 2.2 (0.2) $\times$ 10$^{13}$ \\
    T (K) & - & - & - & - & 55 (8) & 53 (7) & 50 (6) & 55 (7) \\
    \hline
    \multicolumn{9}{c}{CH$_3$OH}\\
    \hline
    261.8057 & $2_{-1,1}-1_{0,1}$ E & 28.0  & -4.2539 & 20 & 10.4 (0.4) & 9.6 (0.4) & 9.1 (0.2) & 10.6 (0.8)  \\
    \hline    
    N (cm$^{-2}$)$^b$ & - & - & - & - & 4.6 (0.7) $\times$ 10$^{15}$ & 4.6 (0.7) $\times$ 10$^{15}$ & 3.6 (0.5) $\times$ 10$^{15}$ & 4.7 (0.6) $\times$ 10$^{15}$ \\
    \hline
    \hline
    \multicolumn{2}{l}{CH$_3$OH/CH$_3$CN} & - & - & - & 119--179 & 117--177 & 151--213 & 177--247 \\
    \hline 
    \end{tabular}
    
\smallskip
{\footnotesize
$^{\rm a}$ Frequencies and spectroscopic parameters are from \citet{cazzoli2006} for CH$_3$CN and \citet{Sastry1984} for CH$_3$OH, extracted from the Cologne Database for Molecular Spectroscopy \citep{Muller2001,Muller2005,Endres2016}.
$^{\rm b}$ Assuming the same temperature as CH$_3$CN.
}
\end{table*}

\section{CH$_3$OH/CH$_3$CN as a potential probe of protostellar conditions}\label{sec:discussion}

Protostellar outflows are time‑dependent structures and the chemistry they host reflects their dynamical evolution. To identify the timescales relevant for our astrochemical modelling, we estimate the dynamical ages of the outflow regions.
We adopt the inclination (70$^\circ$–80$^\circ$ with respect to the line of sight) and inclination-corrected velocities ($\sim$20–50 km s$^{-1}$) derived from $^{12}$CO observations by \citealt{Aso2019}, consistent with previous studies indicating that the outflow lies close to the plane of the sky \citep{Podio2021,LeGouellec2025}.
We compute the dynamical timescales as $t_{\rm dyn}=\frac{d_{\rm true}}{v}=\frac{d_{\rm obs}}{v~\sin(i)}$, where $d_{\rm obs}$ is the projected distance on the plane of the sky, and $i$ and $v$ are the inclination angle and shock velocity, respectively. 
We obtain ages of 300--600 years for R2 and B1, and 500--1000 years for R1 and B2, which are adopted in the following discussion.

\begin{figure}
    \centering
    \includegraphics[width=0.85\linewidth]{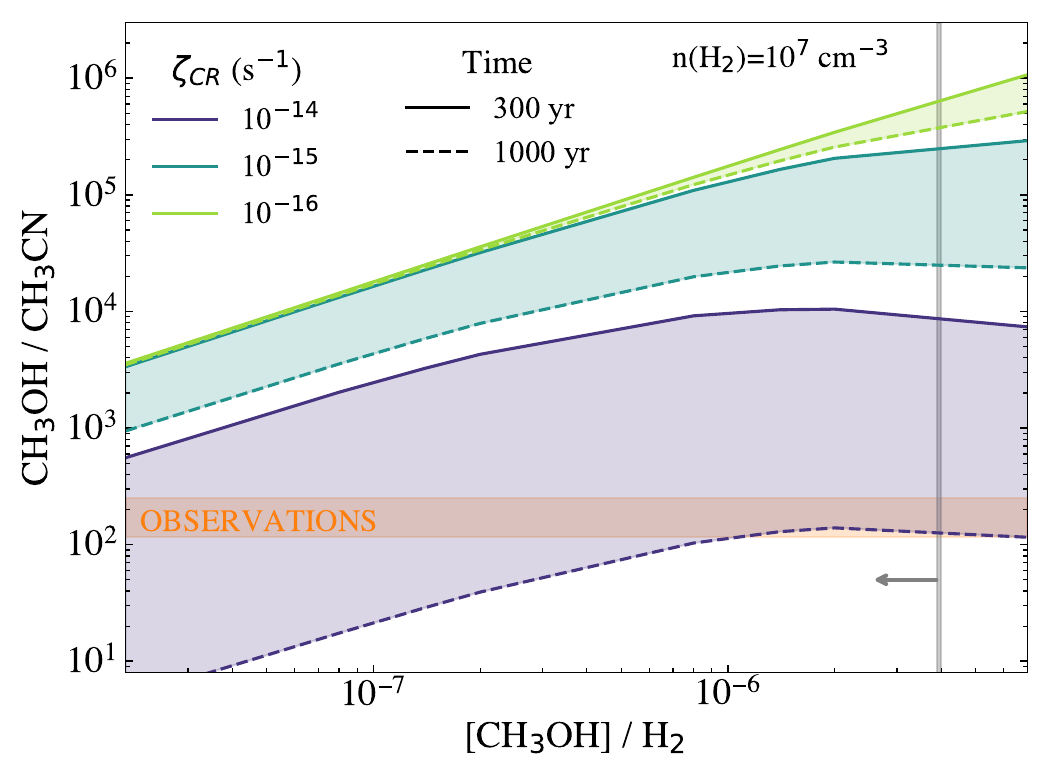}
    \caption{Comparison of predicted and observed CH$_3$OH/CH$_3$CN abundance ratio versus CH$_3$OH/H$_2$ (see Fig.~\ref{fig:models} for other densities). 
    Lines show model predictions at 300 (solid) and 1000 (dashed) yr after shock passage, with shaded areas representing intermediate times.
    Ratios are plotted for $\zeta_{\rm CR}$: 1$\times$10$^{-14}$ (green), 1$\times$10$^{-15}$ (cyan), and 1$\times$10$^{-16}$ (purple) s$^{-1}$. 
    The orange band show the observed ratio (see Tab.~\ref{Tab:lines}), while the vertical grey line marks the CH$_3$OH abundance upper limit (Sec.~\ref{sec:discussion}). 
}
    \label{fig:model}
\end{figure}

To constrain the physical properties of the outflow, we modeled the CH$_3$OH/CH$_3$CN abundance ratio and compared it with the observed values in Tab.~\ref{Tab:lines}. We adopted a pure gas-phase post-shock astrochemical model in which CH$_3$CN is formed in the gas phase through the reactions revised by \cite{giani2023revised} and highlighted in Appendix \ref{app-sec:model}, while methanol is assumed to form on grain surfaces and injected into the gas phase following the passage of the shock. A detailed description of the model is reported in Sect.~\ref{app-sec:model}.
Figure~\ref{fig:model} shows the evolution of the CH$_3$OH/CH$_3$CN ratio as a function of the injected methanol abundance for three different values of the cosmic-ray ionization rate ($\zeta_{\rm CR}$) at different times after the shock passage (between 300 and 1000 yr).
Given the uncertainties in the H$_2$ abundance, and thus in the methanol abundance (Sec.~\ref{sec:results}), we varied the CH$_3$OH abundance between 10$^{-8}$ and 7$\times$10$^{-6}$. The CH$_3$OH/CH$_3$CN ratio is highly sensitive to this parameter as CH$_3$CN precursors can originate either from methanol itself or from other hydrocarbons (see Sec.~\ref{app-sec:model} for details of the reactions).
Since the gas density is not well constrained ($n_{\rm H_2}$>10$^7$ cm$^{-3}$, \citealt{LeGouellec2025} and Appendix~\ref{app-sec:LVG}), we also explored models with different densities (from 10$^6$ to 10$^8$ cm$^{-3}$). 
Within the explored parameter space, the observed CH$_3$OH/CH$_3$CN ratio at the age estimated for the outflow can be reproduced by this gas-phase model only by assuming enhanced values of $\zeta_{\rm CR}$, up to $10^{-14}$ s$^{-1}$.
This result holds even for methanol abundances more than two orders of magnitude below the value derived in Sect.~\ref{sec:results}.

High $\zeta_{\rm CR}$ values have also been inferred in other protostellar environments, such as the outflow regions of OMC-FIR4 and FIR6 \citep{ceccarelli2014-CR, Fontani2017, favre2018-CR, bouvier2025}. Although such high values are not yet fully reproduced by cosmic-ray acceleration models \citep{padovani2016}, previous studies have shown that local particle acceleration can occur in jet-driven shocks, leading to cosmic-ray fluxes significantly higher than the canonical value of $10^{-17}$ s$^{-1}$ \citep{Drury1983,Kirk1994}. Observational evidence suggests that the shock in S68N reaches velocities above 30 km s$^{-1}$ and that the region close to the protostar is strongly magnetized (B$\leq$100 mG) \citep{LeGouellec2025}, making such high $\zeta_{\rm CR}$ values plausible.

Note that the main assumption of our model is that CH$_3$CN is primarily formed in the gas phase, with a negligible contribution from grain-surface chemistry. 
Reproducing the observed ratio with canonical ionization rates would require an additional CH$_3$CN reservoir 2–3 orders of magnitude larger than that predicted by the gas-phase model. 
One possible explanation is that such a reservoir could arise from ice chemistry.
Indeed, alternative formation pathways on dust grains have been proposed, such as the radical–radical reaction CH$_3$ + CN \citep{Hasegawa1993,EnriqueRomero2025} or the hydrogenation of CCN \citep{Garrod2022}.
To reproduce the observations through grain-surface chemistry alone, the abundance of CH$_3$CN in the ice would need to be about a factor 100 lower than that of CH$_3$OH, given the comparable sublimation temperatures of the two species \citep{kakkar2025}. Ice-phase CH$_3$CN remains poorly constrained observationally, with only a tentative detection reported by \citet{Nazari2024}.
If grain-surface formation of CH$_3$CN is indeed efficient, the cosmic-ray ionization rate inferred from our gas-phase model may be overestimated, and the derived values should therefore be regarded as upper limits. A more robust assessment will require chemical models that explicitly include surface reactions, together with the most up-to-date reaction rate coefficients, particularly for CCN hydrogenation, which are currently being investigated (J. Enrique-Romero et al., in preparation; priv. comm.).

While in hot corinos the ratio CH$_3$OH/CH$_3$CN $\sim$100 can be reproduced by pure gas-phase chemistry on timescales of $\sim$10$^4$ yr without invoking enhanced cosmic-ray fields, the situation is different in protostellar outflows. The results of our model suggest that enhanced cosmic-ray ionization rates are required to reproduce CH$_3$OH/CH$_3$CN ratios close to 100 at the typical ages of protostellar outflows ($\sim$300–1000 yr) if only gas-phase chemistry is considered. At such short timescales the CH$_3$OH/CH$_3$CN ratio is highly sensitive to variations in the physical conditions, making it a potentially powerful diagnostic of the environment in shocked gas.
Further work is needed to test the robustness of this method as a probe of the physical conditions in protostellar outflows. First, dedicated models of cosmic-ray acceleration in shocks with the high densities and magnetic fields typical of these regions should be explored. Second, the statistics of CH$_3$OH/CH$_3$CN measurements in outflows should be expanded to determine whether the high cosmic-ray ionization rates inferred here are common or if S68N represents a peculiar case. 

In summary, the S68N outflow shows a  CH$_3$OH/CH$_3$CN ratio of about 100-200, comparable with those found in hot corinos. 
We verified the efficiency of gas phase formation of CH$_3$CN assuming that grain surface production is negligible. We found that an extremely high CR is needed to reproduce the ratio ($\zeta_{\rm CR}$$\sim$$10^{-14}$ s$^{-1}$). 
This shows how powerful the CH$_3$OH/CH$_3$CN correlation can be as probe of the irradiation conditions in protostellar outflows.


\begin{acknowledgements}
We thank the anonymous referee for their comments, which helped improve the manuscript.
The authors thank Prof. C. Ceccarelli and Dr. J. Enrique-Romero for valuable discussions.
LG acknowledges the ESO Scientific Visitor Programme for financial support.
LG, ClCo, LP acknowledge the PRIN-MUR 2020 BEYOND-2p (2020AFB3FX), the project ASI-Astrobiologia 2023 MIGLIORA (F83C23000800005), the INAF-GO 2024 fundings ICES, the INAF-GO 2023 fundings PROTO-SKA (C13C23000770005).
ADZ acknowledge the ESO summer research student program.
This paper makes use of the following ALMA data: ADS/JAO.ALMA\#2017.1.1174.S ALMA is a partnership of ESO (representing its member states), NSF (USA) and NINS (Japan), together with NRC (Canada), NSTC and ASIAA (Taiwan), and KASI (Republic of Korea), in cooperation with the Republic of Chile. The Joint ALMA Observatory is operated by ESO, AUI/NRAO and NAOJ.

\end{acknowledgements}

%
%

\bibliography{LisaGiani}{}
\bibliographystyle{aa}

\begin{appendix} 

\section{Observed spectra and line fits}\label{app-sec:results}

\begin{figure*}
    \centering
    \includegraphics[width=0.9\linewidth]{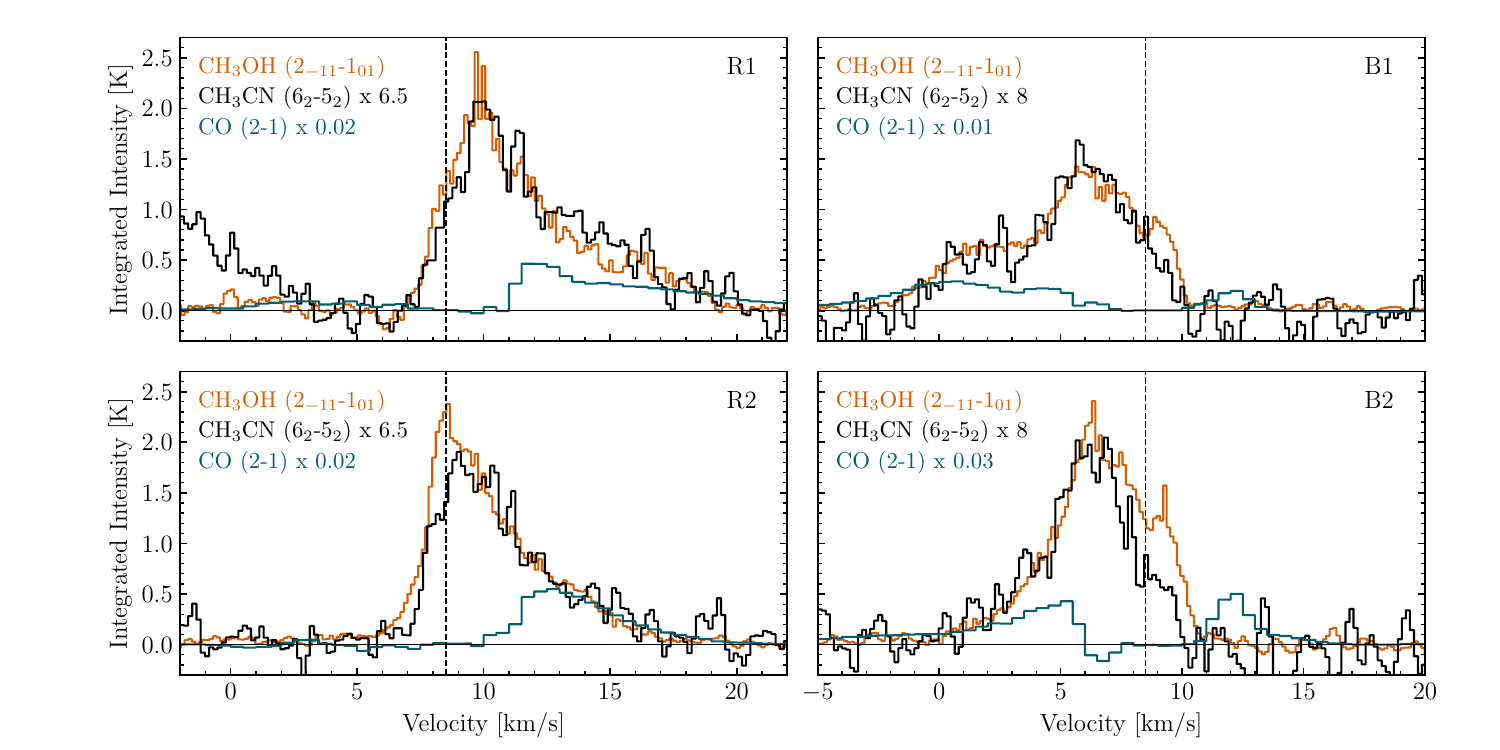}
    \caption{CH$_3$CN 6$_2$-5$_2$ (black), CH$_3$OH 2$_{11}$-1$_{01}$ (orange) and CO 2-1 (teal) spectra (in K) extracted in the four regions of the red-shifted (R1 and R2) and blue-shifted (B1 and B2) outflows of S68N. 
    The spectral resolutions are 0.061 MHz (0.16 km s$^{-1}$) for CH$_3$CN, 0.122 MHz (0.14 km s$^{-1}$) for CH$_3$OH, and 0.384 MHz (0.5 km s$^{-1}$) for CO.
    The CH$_3$CN emission is multiplied by a factor 6.5 in R1 and R2, and by a factor 8 in B1 and B2. The CO emission is multiplied by a factor 0.02 in R1 and R2, 0.01 in B1 and 0.03 in B2 for visualization purposes. The black dashed vertical lines mark the v$_{\rm sys}$ (+8.5 km s$^{-1}$, \citealt{Lee2014}). }
    \label{fig:spectra-ch3oh-ch3cn-CO}
\end{figure*}

\begin{figure*}
    \centering
    \includegraphics[width=0.98\linewidth]{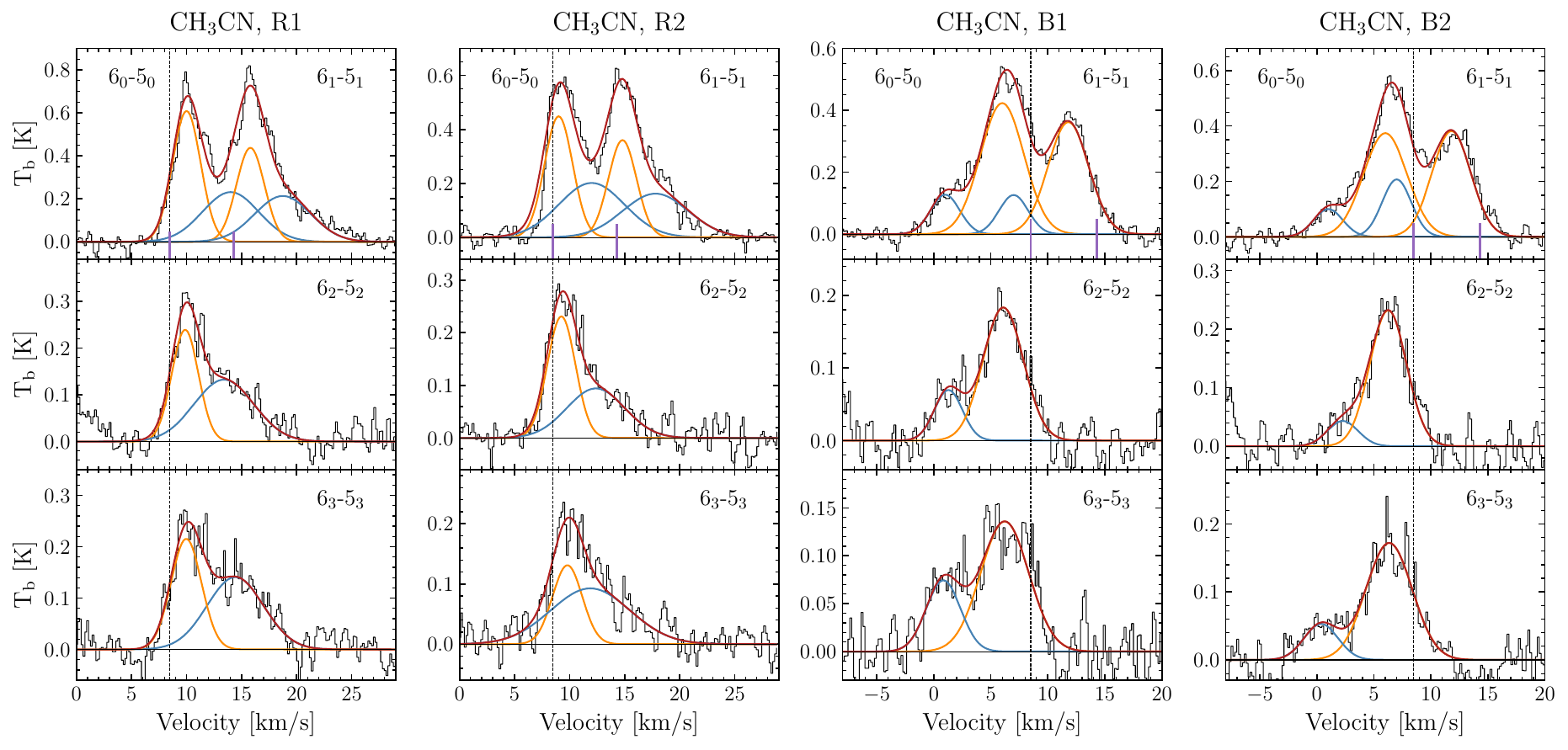}
    \caption{Spectral fits of the CH$_3$CN 6$_0$–5$_0$, 6$_1$–5$_1$, 6$_2$–5$_2$, and 6$_3$–5$_3$ transitions extracted in the four regions R1, R2, B1, and B2.
    In the upper panels of each region, the transitions producing the emission lines (see Table~\ref{Tab:lines}) are indicated, and their corresponding frequencies are marked by small vertical purple lines. The spectra are centred at the frequency of the 6$_0$–5$_0$ transition (110.3835 GHz).
    In regions R1 and R2, each transition shows a main peak at +9–10 km s$^{-1}$ (orange curves) and a secondary component redshifted by $\sim$3–4 km s$^{-1}$ (blue curves).
    In regions B1 and B2, each transition shows a main peak at +6 km s$^{-1}$ (orange curves) and a secondary component blueshifted by $\sim$5 km s$^{-1}$ (blue curves).
    In all panels, the red curves show the total multi-component fits. The dashed vertical line marks the systemic velocity, +8.5 km s$^{-1}$ (\citealt{Lee2014}).
    }
    \label{fig:spectra-ch3cn-deblending}
\end{figure*}

\begin{figure*}
    \centering
    \includegraphics[width=0.95\linewidth]{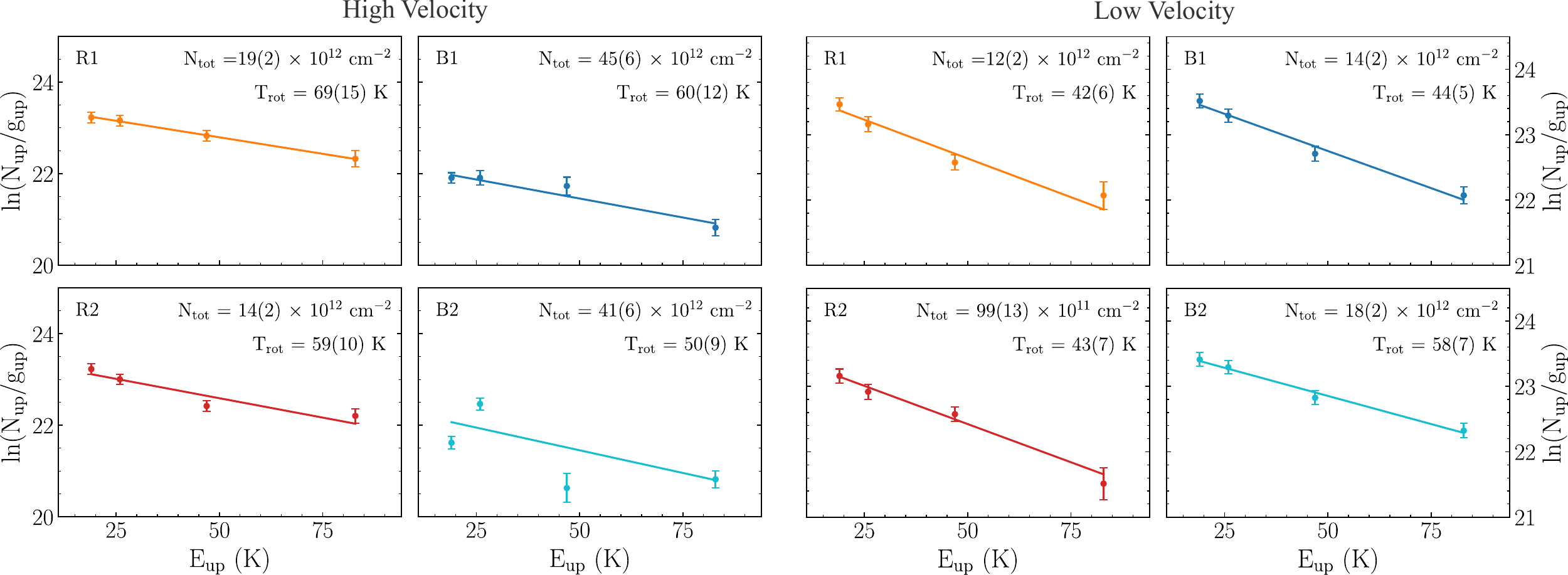}
    \caption{
    CH$_3$CN rotation diagram with fit for the four regions R1, R2, B1 and B2 of the S68N outflows obtained separating the high and low velocity components of the fits. The color coding is the same used in Fig.~\ref{fig:maps+RD} to identify the four regions. The resulting column densities and rotational temperatures are reported on the top right corner of each diagram. 
    }
    \label{fig:RD-high-low}
\end{figure*}

Figures~\ref{fig:spectra-ch3oh-ch3cn-CO} and \ref{fig:spectra-ch3cn-deblending} show the CH$_3$OH, CH$_3$CN, and CO spectra extracted in the four regions of the S68N outflow, together with the multi-component fits used to analyse the CH$_3$CN emission.  
To verify if the gas properties change at different velocities, we performed a separate analysis for the high- and low-velocity kinematic components, shown in the rotation diagrams of Figure~\ref{fig:RD-high-low}. 
Table~\ref{Tab:lines-high-low} reports the results for both regimes, including column densities, temperatures, and abundance ratios.
We find no significant variations in the column densities of CH$_3$CN and CH$_3$OH with increasing distance from the protostar. However, the rotational temperatures derived from the high-velocity components are higher than those of the low-velocity gas, likely reflecting the more efficient heating and compression produced by shocks in the fastest-moving material. 
The CH$_3$OH/CH$_3$CN abundance ratios for both velocity components are consistent, within uncertainties, with those derived from the combined (high+low) velocity analysis. This suggests that, despite the different thermal conditions, the relative chemical abundances remain robust and do not show significant variations along the outflow or between different velocity regimes.

\renewcommand{\arraystretch}{1.2}
\begin{table*}[]
    \begin{center}
    \caption{Line fit results of CH$_3$CN and CH$_3$OH from the high and low velocity components (see Sec.~\ref{sec:results}). \label{Tab:lines-high-low}}
    \begin{tabular}{c|cccc|cccc}
    \hline
    \hline
    \multicolumn{1}{c}{} & \multicolumn{4}{c}{Low Velocity} & \multicolumn{4}{c}{High Velocity} \\
    \hline    
    \multicolumn{1}{c}{} &  \multicolumn{8}{c}{CH$_3$CN} \\
    \hline
    Frequency & I$_{\rm int}$ [R1] & I$_{\rm int}$ [R2] & I$_{\rm int}$ [B1] & I$_{\rm int}$ [B2] & I$_{\rm int}$ [R1] & I$_{\rm int}$ [R2] & I$_{\rm int}$ [B1] & I$_{\rm int}$ [B2]\\
    (GHz) & & (K)  & (s$^{-1}$) & & (K\ km\ s$^{-1}$) & (K\ km\ s$^{-1}$)  & (K\ km\ s$^{-1}$) & (K\ km\  s$^{-1}$) \\ 
    \hline
    110.3835 & 1.95 (0.04) & 1.43 (0.05) & 2.03 (0.06) & 1.79 (0.06) & 1.48 (0.09) & 1.50 (0.10) & 0.41 (0.02) & 0.32 (0.03) \\
    110.3814 & 1.40 (0.08) & 1.15 (0.06) & 1.61 (0.03) & 1.63 (0.03) & 1.36 (0.08) & 1.21 (0.06) & 0.40 (0.05) & 0.66 (0.06) \\
    110.3750 & 0.73 (0.04) & 0.75 (0.03) & 0.81 (0.04) & 0.94 (0.03) & 0.90 (0.06) & 0.62 (0.04) & 0.28 (0.05) & 0.15 (0.03) \\
    110.3644 & 0.75 (0.12) & 0.45 (0.09) & 0.72 (0.06) & 0.87 (0.04) & 0.91 (0.13) & 0.84 (0.10) & 0.22 (0.03) & 0.20 (0.03) \\
%
    \hline
    N (cm$^{-2}$) $\times$ 10$^{13}$ & 1.2 (0.2) & 0.99 (0.13) & 1.4 (0.2) & 1.8 (0.2) & 1.9 (0.2) & 1.4 (0.2) & 0.45 (0.06) & 0.41 (0.06) \\
    T (K)       & 42 (6) & 43 (7) & 44 (5) & 58 (7) & 69 (15) & 59 (10) & 60 (12) & 50 (9) \\
    \hline
    \multicolumn{1}{c}{} &  \multicolumn{8}{c}{CH$_3$OH} \\
    \hline
    261.8057 & 5.5 (0.2) & 5.29 (0.19) & 6.8 (0.2) & 7.5 (0.6) & 4.9 (0.3) & 4.6 (0.2) & 2.35 (0.12) &  3.0 (0.6) \\ 
    N (cm$^{-2}$) $\times$ 10$^{15}$  & 1.8 (0.3) & 1.8 (0.3) & 2.3 (0.3) & 3.5 (0.5) & 2.9 (0.8) & 2.2 (0.4) & 1.1 (0.2) & 1.2 (0.3) \\
    \hline
    \multicolumn{1}{c}{} &  \multicolumn{8}{c}{CH$_3$OH/CH$_3$CN} \\
    \hline
     & 115-185 & 140-210 & 132-196 & 159-229 & 100-200 & 120-200 & 180-300 & 210-370 \\
    \hline
    \end{tabular}
    \end{center}
\end{table*}

\section{LVG analysis}\label{app-sec:LVG}

We used a non-LTE analysis via our in-home large velocity gradient (LVG) code \texttt{grelvg} \citep{Ceccarelli2002,Ceccarelli2003} to predict the molecular line intensities that are simultaneously fitted via comparison to the observed ones using a $\chi^2$ minimization.
The collisional coefficients of CH$_3$CN are reported in the LAMDA database$^1$, computed by \citep{BenKhalifa2023} between 20 and 100 K for the lowest 50 levels of A-CH$_3$CN-He (ortho) and E-CH$_3$CN-He (para) and scaled for collisions with H$_2$. We assumed a semi-infinite expanding slab geometry, the H$_2$ ortho-to-para ratio equal to 3 and the CH$_3$CN A-to-E ratio equal to 1. 

\begin{figure*}
    \centering
    \includegraphics[width=0.8\linewidth]{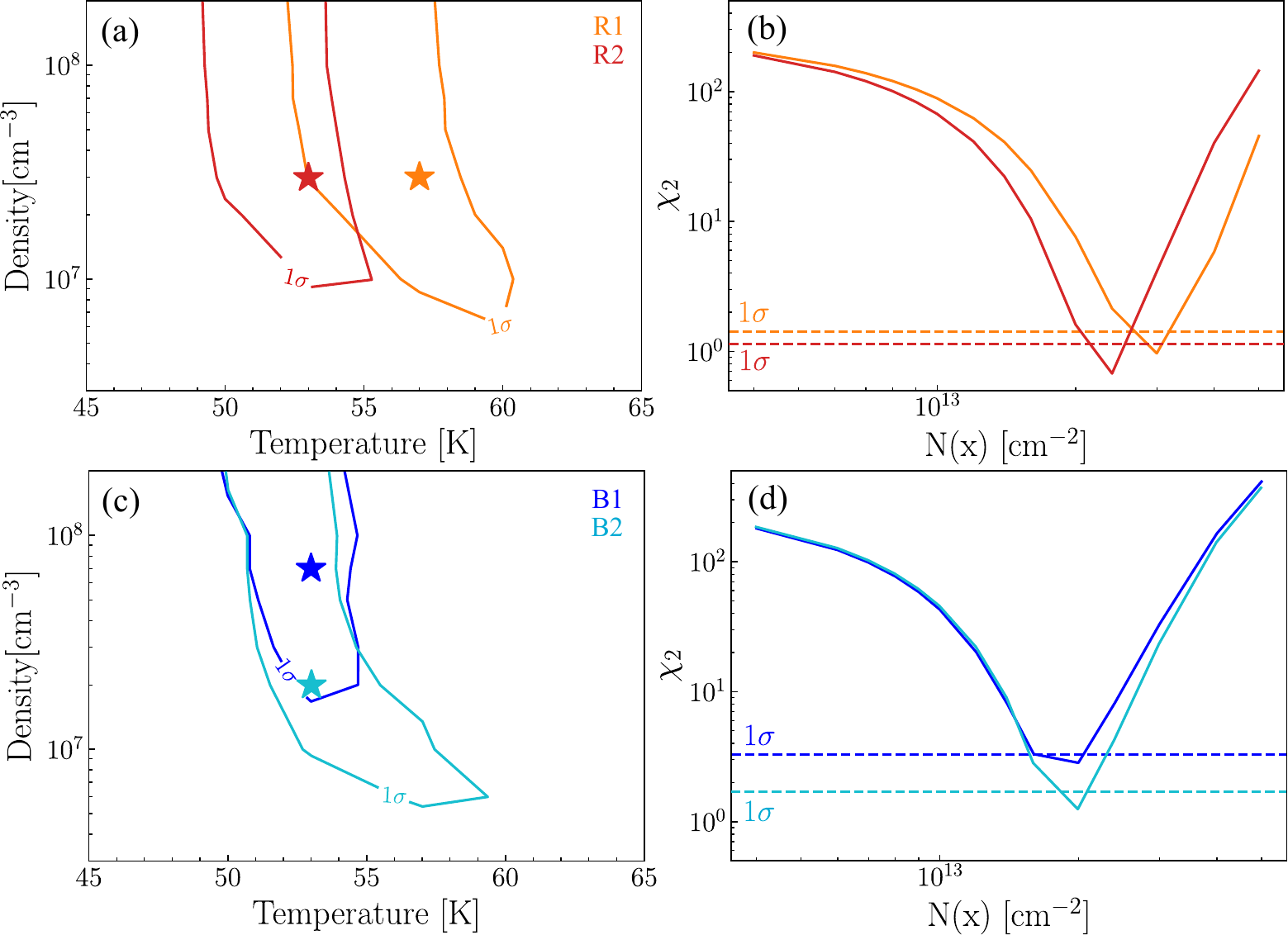}
    \caption{LVG analysis of CH$_3$CN in the S68N outflows. 
    \textit{Panels (a) and (b)}: Density--temperature $\chi^2$ contour plot (a) and reduced $\chi^2$ versus $N_{\text{CH}_3\text{CN}}$ plot (b) for R1 (orange) and R2 (red). In panel (a), the contours represent the 1$\sigma$ confidence levels, and the best-fit solutions for R1 and R2 are marked by orange and red stars, respectively. Color coding matches that of Fig.~\ref{fig:maps+RD}.
    \textit{Panels (c) and (d)}: Same as (a) and (b), but for the blue-shifted regions B1 (blue) and B2 (cyan).}
    \label{fig:lvg}
\end{figure*}

We ran a large grid of models covering the frequency of the observed CH$_3$CN lines, with a total (A plus E) column density $N_{\text{CH}_3\text{CN}}$ ranging from $4 \times 10^{12}$ to $6 \times 10^{13} \, \text{cm}^{-2}$, a gas density $n_{\text{H}_2}$ from $3 \times 10^6$ to 10$^9$ $\text{cm}^{-3}$, and a temperature $T$ from 40 to 85 K. 
We simultaneously fit the measured CH$_3$CN line intensities in R1, R2, B1 and B2 via comparison with those simulated by the LVG model, leaving $N_{\text{CH}_3\text{CN}}$, $n_{\text{H}_2}$ and  $T$, and assuming that the source is extended. 
Following the observations, we assumed a line width equal to $5 \, \text{km s}^{-1}$ and 
included the calibration uncertainty ($10\%$) in the observed intensities. 
The best-fit parameters obtained from our LVG analysis are summarized in Tab.~\ref{tab:lvg} and Fig.~\ref{fig:lvg}. The obtained densities ($n_{\text{H}_2}$$\geq$ $5.5 \times 10^6$ cm$^{-3}$) and optical depth values ($\tau\leq$0.01) means LTE and optically thin conditions are valid for CH$_3$CN.
Indeed, the resulting excitation temperatures and CH$_3$CN column densities are in agreement with the values derived through the RD method (see Table~\ref{Tab:lines-high-low}).
Furthermore, the gas densities derived for all regions are $\gtrsim$10$^7$ cm$^{-3}$, which is consistent with the high-density environment of the S68N outflows previously reported by \citet{LeGouellec2025} in the region close to the protostar.

\begin{table*}[]
\renewcommand{\arraystretch}{1.3}
\centering
    \caption{
    Best-fit results and 1$\sigma$ (50\%) confidence level (range) from the non-LTE LVG analysis of the CH$_3$CN lines.
    }
    \label{tab:lvg}
    \begin{tabular}{lcccccccc}
    \hline
    \hline
    \multicolumn{1}{c}{} & \multicolumn{2}{c}{R1} & \multicolumn{2}{c}{R2} & \multicolumn{2}{c}{B1} & \multicolumn{2}{c}{B2} \\
    \cline{2-3} \cline{4-5} \cline{6-7} \cline{8-9}
    \multicolumn{1}{c}{} & Best Fit & Range & Best Fit & Range & Best Fit & Range & Best Fit & Range \\
    \hline																					
    $n_{\rm H_2}$ (cm$^{-3}$) & $3.0 \times 10^{7}$ & 	$ \ge 7.0 \times 10^{6}$ & $3.0 \times 10^{7}$ &  $ \ge 9.0 \times 10^{6}$ & $7.0 \times 10^{7}$ & $ \ge 1.5 \times 10^{7}$ & $2.0 \times 10^{7}$ & $ \ge 5.5 \times 10^{6}$  \\		
    $T_{\rm kin}$ (K) & 57 & 53-60 & 53 &  49-55 & 53 & 49-55 & 53 & 50-59  \\	
    $N_{\rm CH_3CN}$  & $3.0 \times 10^{13}$ & - & $2.4 \times 10^{13}$ & - & $2.0 \times 10^{13}$ & - & $2.0 \times 10^{7}$ & -  \\
    \hline
    \end{tabular}
\end{table*}

\section{Model description}\label{app-sec:model}

To estimate the abundances of CH$_3$OH and CH$_3$CN, we employed the \texttt{MyNahoon} chemical code \citep{wakelam2005estimation,wakelam2010sensitivity}, which calculates gas–phase abundances for a given set of physical parameters (temperature $T$, density $n_{\rm H_2}$, visual extinction $A_{\rm v}$, and cosmic ray ionization rate $\zeta_{\rm CR}$). The simulations make use of the GRETOBAPE gas-phase chemical network \citep{tinacci2023-gretobape}, which incorporates several updated reactions based on studies carried out by our group and others \citep{loison2014,balucani2015formation,skouteris2018genealogical,vazart2020gas,blazquez2020gas,giani2023revised,Giani2025}.

To investigate how the passage of a shock affects the chemical composition of the gas, we adopted a two-step approach similar to that previously applied to model protostellar molecular shocks \citep[e.g.,][]{Podio2014,codella2017,desimone2020,bouvier2025}.
In the first step, we derived the chemical composition of the pre-shock cloud starting from the elemental abundances listed in Tab. \ref{tab:model-initial+injected}. The physical conditions were set to $T$ = 10 K, $n_{\rm H_2} = 1\times10^4$ cm$^{-3}$, $A_{\rm v} = 100$ mag, and a cosmic-ray ionization rate $\zeta_{\rm CR} = 1\times10^{-17}$ s$^{-1}$.
In the second step, we simulated the effect of mantle sputtering induced by the shock by enhancing the abundances of several species released from dust grain mantles (the injected species and their abundances are reported in Tab. \ref{tab:model-initial+injected}). 
The injected species are those generally considered to form predominantly on grain surfaces and for which gas-phase formation is inefficient. The formation of CH$_3$CN, however, is debated (see Sec.~\ref{sec:discussion}). For this reason, we do not inject CH$_3$CN into the gas phase, as our goal is to assess the efficiency of its gas-phase formation under the assumption of a negligible grain-surface contribution. 
We then followed the chemical evolution under post-shock conditions assuming a gas temperature of $\sim$60~K (derived from CH$_3$CN observations, see Sec.~\ref{sec:results}) and densities in the range $10^6$–$10^8$ cm$^{-3}$.
We also tested temperatures between $\sim$40 and 70 K, consistent with the values reported in Table~\ref{Tab:lines}. The CH$_3$OH/CH$_3$CN ratio is only weakly affected by the adopted temperature, varying by less than 10\% over this range.

We adopted different values of the cosmic-ray ionization rate, namely $\zeta_{\rm CR}=10^{-16}$, $10^{-15}$, and $10^{-14}$ s$^{-1}$. Due to the high densities considered, the outflow region was assumed to be highly shielded from external radiation, and therefore a visual extinction of $A_{\rm v}=100$ was adopted.
The abundance of injected methanol was treated as a free parameter and varied between $1\times10^{-8}$ and $7\times10^{-6}$, in agreement with the upper limit derived in Sec.~\ref{sec:discussion} ($\lesssim$5$\times$10$^{-6}$).

\begin{figure*}
    \centering
    \includegraphics[width=0.98\linewidth]{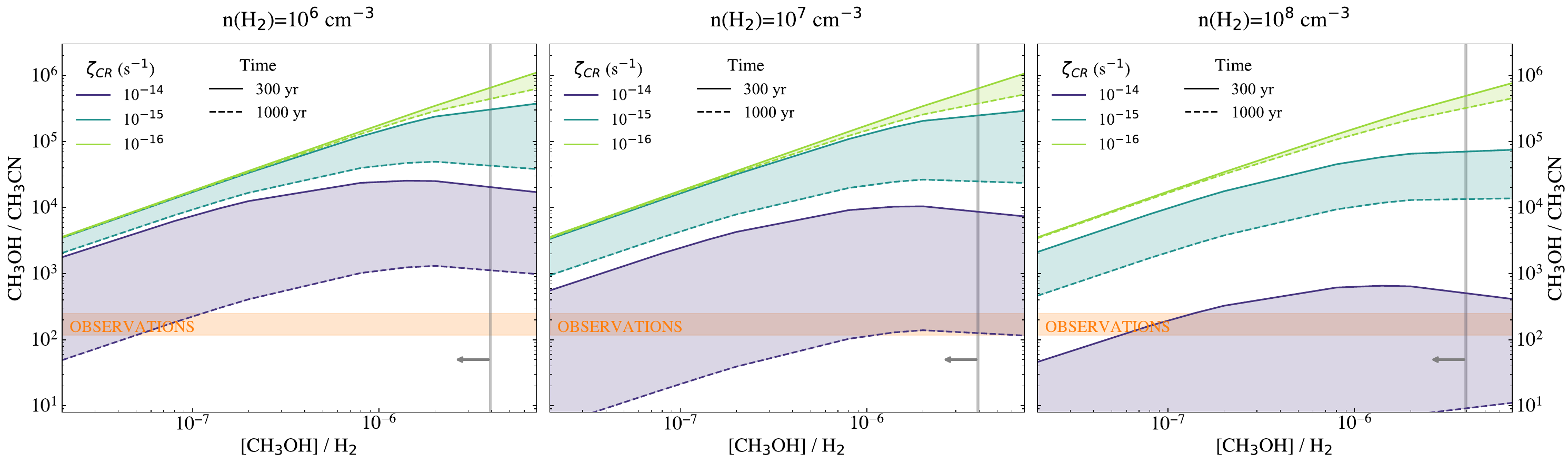}
    \caption{Same of Fig.~\ref{fig:model} but for densities of 10$^6$ (left panel), 10$^7$ (middle panel) and 10$^8$ (right panel) cm$^{-3}$.
    }
    \label{fig:models}
\end{figure*}

The model results are shown in Fig.~\ref{fig:models}.
In general, higher densities lead to faster chemical evolution, resulting in lower CH$_3$OH/CH$_3$CN abundance ratios at earlier times. 
An increase in the cosmic-ray ionization rate ($\zeta_{\rm CR}$) also leads to a significant decrease in the CH$_3$OH/CH$_3$CN abundance ratio. 
The main formation pathway of methyl cyanide in outflows involves two steps: the radiative association of CH$_3^+$ with HCN to form protonated methyl cyanide (CH$_3$CNH$^+$), followed by proton transfer to ammonia (CH$_3$CNH$^+$ + NH$_3$) producing CH$_3$CN.
At low ionization rates ($\zeta_{\rm CR} \sim 10^{-16}$ s$^{-1}$), methanol is mainly destroyed through reactions with H$_3^+$, enhancing the production of CH$_3^+$, the key precursor in the formation of CH$_3$CN.
At higher ionization rates ($\zeta_{\rm CR} \sim 10^{-14}$ s$^{-1}$), the chemistry changes: methanol destruction is dominated by reactions with H$_3$O$^+$, while CH$_3^+$ is formed via the CH$_2^+$ + H$_2$ reaction. The combined effect of more efficient methanol destruction and enhanced CH$_3$CN production results in a marked reduction of the CH$_3$OH/CH$_3$CN ratio.

\begin{table*}[]
\renewcommand{\arraystretch}{1.3}
\centering
    \caption{Initial elemental abundances relative to H nuclei adopted for the cold molecular cloud model \citep{jenkins2009unified} (left columns) and abundances of species injected into the gas phase after the shock passage (right columns).}
    \label{tab:model-initial+injected}
    \begin{tabular}{cccc|cccc}
    \hline
    \hline
    \multicolumn{4}{c|}{Initial abundances} &  \multicolumn{4}{c}{Injected abundances} \\
    \hline
    Element  & Abundance &  Species & Abundance & Element  & Abundance &  Species & Abundance\\
    \hline																						
    He      	& 	$9.0 	\times 	10^{-2}$  &  P$^+$  & 	$2.0\times10^{-10}$ & 	H$_{2}$O	& 	$1\times10^{-4}$     & 	OCS     	     & $2\times 10^{-6}$  \\		
    C$^+$   	& 	$1.7 	\times 	10^{-5}$  &  Na$^+$ & 	$2.0\times10^{-9}$  & 	CO$_{2}$	& 	$3\times10^{-5}$     & 	SiO     	     & $1\times 10^{-6}$  \\		
    O       	& 	$2.6 	\times 	10^{-5}$  &  Mg$^+$ & 	$7.0\times10^{-9}$  & 	CO          & 	$1\times10^{-4}$     & 	Si      	     & $1\times10^{-6}$   \\		
    N       	& 	$6.2 	\times 	10^{-6}$  &  Fe$^+$ & 	$3.0\times10^{-9}$  & 	CH$_3$OH	& $0.1-70\times 10^{-7}$ & 	CH$_3$CH$_{2}$	 & $4\times 10^{-8}$  \\
    S$^+$   	& 	$8.0 	\times 	10^{-8}$  &  Cl$^+$ & 	$1.0\times10^{-9}$  & 	NH$_3$  	& $5.6\times10^{-5}$     & 	CH$_3$CH$_{2}$OH & $6\times 10^{-8}	$ \\
    Si$^+$  	& 	$8.0 	\times 	10^{-9}$  &  F$^+$  & 	$1.0\times10^{-9}$  & 	H$_{2}$CO   & 	$1\times10^{-6}$     & 	SiH$_4$          & $1\times10^{-7}$   \\		
    \hline
    \end{tabular}
\end{table*}

\end{appendix}

\end{document}